\documentclass[twocolumn]{emulateapj}
\submitted{Science with Parkes @ 50 Years Young, 31 Oct. -- 4 Nov., 2011}

\shorttitle{Planning and Construction
of the Parkes Telescope}
\shortauthors{Peter Robertson}

\begin{document}

\title{An Australian Icon -- Planning and Construction
of the Parkes Telescope}

\author{Peter Robertson}

\affil{School of Physics, University of Melbourne, Vic. 3010, Australia}
\email{prob@unimelb.edu.au}

\begin{abstract}

By almost any measure, the Parkes Radio Telescope is the most successful scientific instrument ever built in Australia.  The telescope is unsurpassed in terms of the number of astronomers, both national and international, who have used the instrument, the number of research papers that have flowed from their research, and the sheer longevity of its operation (now over fifty years).  The original planners and builders could not have envisaged that the telescope would have such an extraordinarily long and productive future.  From the start, it was an international project by CSIRO that in the 1950s launched Australia into the world of `big science'.  Partly funded by the US Carnegie and Rockefeller foundations, it was designed in England by Freeman Fox \& Partners, and built by the German firm MAN.  This article will give an overview of the origins of the idea for the telescope and the funding, planning and construction of the Parkes dish over the period 1954 to 1961.
\end{abstract}


\section{Introduction}

The construction of the Parkes Telescope brought what has been termed `Big Science' to Australia.  Until then no university department or government research lab in Australia had anywhere near the resources to undertake such an expensive and technologically advanced project.  The one possible exception was the construction of the 74-inch (1.9 m) telescope at Mt Stromlo near Canberra in 1955, making it the equal largest optical telescope in the southern hemisphere.  Funded by the Commonwealth government, the Mt Stromlo telescope was based on a `tried and true' technology, whereas the Parkes Telescope was to be a leap into an engineering unknown.

From the outset the Parkes Telescope was an international project.  Approximately half the funds came from American philanthropic foundations, the design was carried out by a British engineering firm and, after an international tender, the construction contracted to a German company.  In this article we will trace how the decision was made to build a Giant Radio Telescope (GRT) in Australia; how the instrument was to be funded (1954--55); the planning and design of the telescope (1956--59); the search for a suitable site (1958); and conclude with an overview of the construction phase (1959--61). Most of this article is based on a previous detailed study on the history of the Parkes Telescope (Robertson 1992).

\section{Why build a Giant Radio Telescope?}

The origins and early development of radio astronomy in Australia is possibly the most intensively studied subject in the history of Australian science.  Over the past decade or so a number of prominent historians of astronomy have carried out detailed studies of various aspects of early Australian radio astronomy -- see for example Orchiston (2005), Sullivan (2009) and Goss \& McGee (2010).  For the first time in its history -- at least in the physical sciences -- Australia pioneered and became a world leader in a major new field of science.

To understand how radio astronomy started in Australia, it is necessary to go back to the beginning of World War II.  At the time the Australian government learnt that the British had developed a new secret weapon known as radar (originally known as radio direction finding).  Radar would become of vital importance and a major reason why the Allies went on to win the war.  The Australian government decided to set up a secret laboratory which could carry out research and development of radar equipment suitable for use in the Pacific region.  The laboratory would be a division within Australia's leading research organisation, the Council for Scientific \& Industrial Research [the forerunner of the Commonwealth Scientific \& Industrial Research Organisation (CSIRO)].  The new division was given the innocuous title of the Radiophysics Laboratory to hide its true purpose and a special building was rapidly built for it in the grounds of the University of Sydney.

Towards the end of the war, when wartime radar research was no longer required, CSIRO needed to make a very important decision.  Should it disband the Radiophysics Lab and send its staff back to their peacetime professions?  This is in fact what happened to similar radar labs in Britain and the United States.  Instead, the CSIRO decided to keep the Lab in tact and redirect its research into peaceful applications of the new radar technology.  In 1945 a long list of possible applications was drawn up, including using radar to improve air navigation and to study the physical processes of rain formation in clouds.

Another project was to investigate reports by a number of wartime radar operators that the Sun is a strong emitter of radio waves.  A related project was an investigation  the reports by Karl Jansky and Grote Reber in the United States of radio emission from the Milky Way galaxy, a new and mysterious phenomenon.  These related projects turned out to be the wild card in the pack and brought almost instant success.  By about 1950 radio astronomy -- as it became known -- had grown to be the main research area of the Lab.  Together with the two groups in Britain at Jodrell Bank (University of Manchester) and at the Cavendish Laboratory (Cambridge University), the Radiophysics Lab had become a world leader in a brand new branch of astronomy.

\begin{figure}
\resizebox{\hsize}{!}{\includegraphics{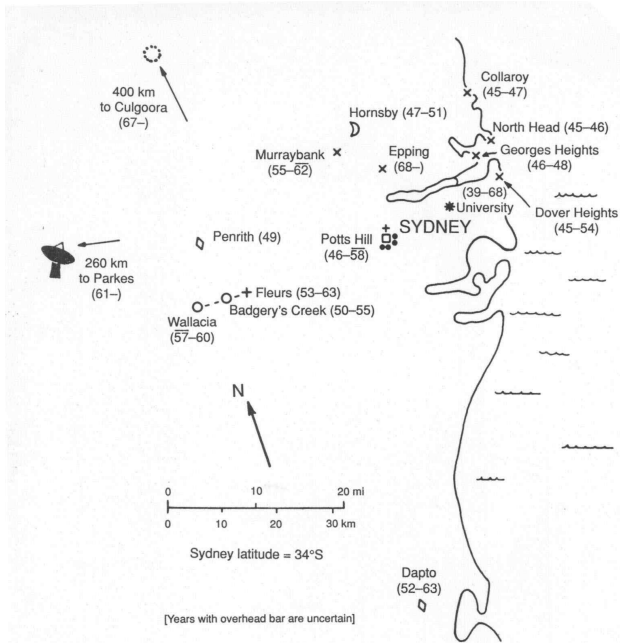}}
\caption{Location of the field stations in and around Sydney used by the Radiophysics Lab after the end of WWII [courtesy: W. T. Sullivan].}
\end{figure}

The Lab developed a wide variety of radio telescopes to study solar radio emission and `cosmic noise' from the galaxy as it was then known.  These telescopes were located at a number of field stations in and around Sydney, chosen because of their geographical advantages and because they were relatively free of manmade sources of radio interference (see Fig. 1).  There were only a handful of scientific and technical staff at each site and each group was largely autonomous.  Most staff would go into the Lab only occasionally -- perhaps to have some equipment made in the workshops or to attend a staff meeting.  The `glue' that held these semi-independent groups together was Joe Pawsey, who was head of the radio astronomy group and assistant chief of the Lab.  Pawsey would regularly visit each field station to discuss any problems and to pass on news of progress elsewhere in the Lab.

By about 1950 a debate arose within the group about where this rapid growth in radio astronomy was heading.  Should the Lab continue with a number of small groups, each working independently, or should it begin to focus on one or two large projects that would bring these small groups together?  Joe Pawsey and a number of senior staff were in favour of continuing with the status quo, an approach that had brought world acclaim to the Lab.  Why make changes to something that was working so well?  This approach was entirely consistent with Pawsey's own background.  Educated at the University of Melbourne, he went on to complete a PhD at the Cavendish on radio propagation in the ionosphere.  He became an advocate of the so-called `string and sealing wax' approach to research at the Cavendish, where innovative solutions to problems could found using simple and inexpensive equipment.

\begin{figure}
\resizebox{\hsize}{!}{\includegraphics{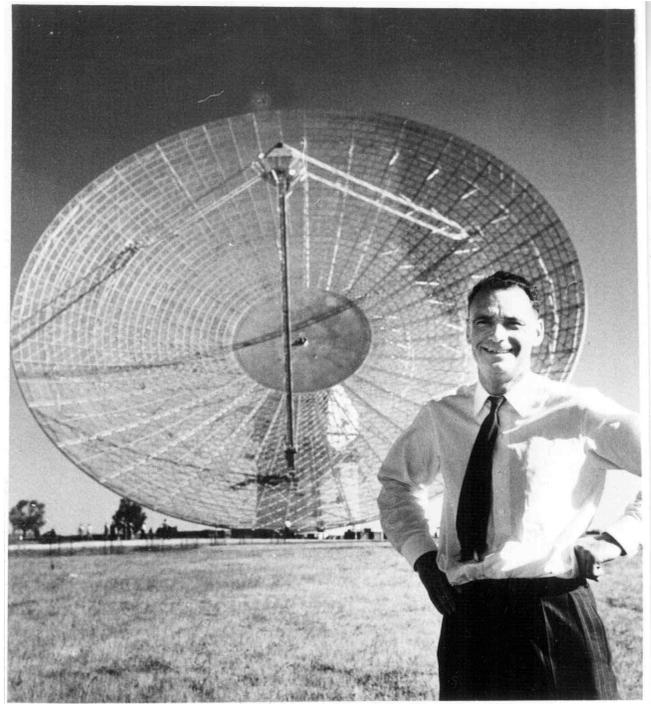}}
\caption{Taffy Bowen was the driving force behind plans to build a Giant Radio Telescope in Australia. He was Chief of the Radiophysics Lab from 1946 until his retirement from CSIRO in 1971.}
\end{figure}

This was not the view, however, of E. G. `Taffy' Bowen who was Chief of the Radiophysics Lab and to whom Pawsey reported (see Fig. 2).  Bowen, and others such as John Bolton (the inaugural director of the Parkes Telescope), believed that radio astronomy would soon develop in the same way as conventional astronomy, where the best science and the important discoveries were carried out with the largest and most powerful optical telescopes.  As his name suggests, Taffy Bowen was born and educated in Wales and completed a PhD at King's College, London in physics.  In 1935 he joined the English group secretly developing the first radar equipment.  At this time radar equipment was large and cumbersome and was suitable only for land- or ship-based applications.  Bowen's great achievement was to miniaturise the equipment so that it could fit into the nose of a plane -- and so created the first airborne radar (Bowen 1987).

The event that convinced Bowen that a large general purpose telescope was the way of the future came in 1951 when Bernard Lovell's group at Jodrell Bank announced plans to build a giant parabolic dish (see Fig. 3).  Bowen feared that the new telescope would scoop the pool of discoveries and that the Radiophysics Lab would soon lose its position as a world leader in radio astronomy.  Bowen believed that to avoid being left behind, Radiophysics would have to build an instrument at least as good as, if not better than, the Jodrell giant.  Although the approach of small semi-autonomous groups at Radiophysics continued to the late 1950s, Bowen's vision of a Giant Radio Telescope eventually prevailed.

\begin{figure}
\resizebox{\hsize}{!}{\includegraphics{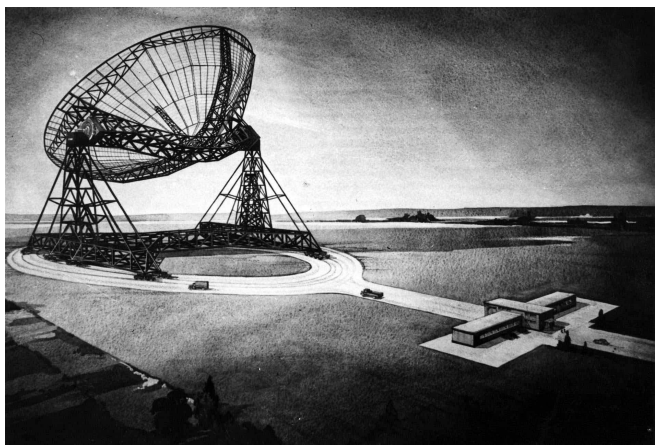}}
\caption{In 1951 the Jodrell Bank group in Manchester announced plans to build a giant dish with a diameter of 250 ft (76 m).}
\end{figure}

\section{Finding the Funds}

The immediate problem Bowen faced was where would the money come from to build a GRT?  Radio astronomy was only one of several research projects in the Lab and as chief he had to work within a fixed annual budget.  Similarly, the Radiophysics Lab was only one of a dozen divisions within CSIRO, all clamouring for more funding to support their expanding postwar research programs.  

At Jodrell Bank, Bernard Lovell had received a large grant from the philanthropic Nuffield Foundation to partly fund his large telescope. Lovell had then persuaded the UK's Department of Scientific \& Industrial Research to fund the remainder of the projected costs.  Bowen realised he would need to explore a similar path, but was faced with the reality that there were no comparable philanthropic foundations in Australia to support scientific research, especially in a new and obscure area such as radio astronomy. 

After his pioneering work on airborne radar, Bowen had spent most of the war years shuttling back and forth between England and the United States as the principal liaison officer coordinating the development of radar in the two countries. With England's industrial capacity to produce conventional weapons stretched to a breaking point, it had decided to rely on America's industrial might to develop and mass produce a variety of radar equipment which would prove so decisive to the outcome of the war.

\begin{figure}
\resizebox{\hsize}{!}{\includegraphics{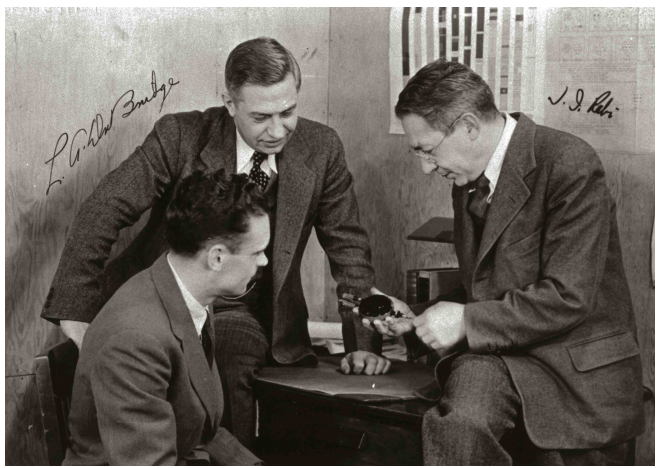}}
\caption{Taffy Bowen at the MIT Rad Lab with Lee Dubridge, who was appointed president of the California Institute of Technology (Caltech) in 1946, and Isidor Rabi who was awarded the Nobel Prize for Physics in 1944 [courtesy: Bowen family].}
\end{figure}

In the United States, Bowen spent most of his time at the Massachusetts Institute of Technology in Boston, which had established the Radiation Laboratory to coordinate American research and development of radar -- the US equivalent of the Radiophysics Lab but on a much bigger scale. While at the MIT Rad Lab, as it was known, Bowen got to know many powerful men in American science, including some who had considerable influence within the network of large US philanthropic organisations (see Fig. 4).  He would draw on this `old-boy network' to provide the funds to build a GRT in Australia.

The breakthrough occurred when Bowen learnt that the Carnegie Corporation administered what was known as the British Dominion and Colonies Fund, which had been established specifically to support projects within Britain and her Commonwealth countries.  The Fund had been suspended at the beginning of the war and its capital had been steadily growing.  In August 1952 Bowen wrote to Vannevar Bush, president of the Carnegie Institution in Washington, and enquired about the possibility of funding for an Australian GRT.  He was informed that the trustees of the Fund had decided that, rather than support a range of small projects, it wanted to make one single large grant.  Bowen was in the right place at the right time.  In May 1954, the Carnegie Corporation announced that it would provide \$US 250,000 (\pounds A 112,000) towards the partial funding of a GRT in Australia.

With the Carnegie promise, Bowen's next challenge was to seek funding from the Australian government.  With the support of CSIRO head office, the request went to Richard Casey, the Minister-in-Charge of CSIRO (see Fig. 5).  Casey had developed a personal interest in the Radiophysics achievements and had no hesitation in discussing the GRT project with prime minister Robert Menzies.  Although he agreed to match the Carnegie grant with government funds, Menzies insisted that at least one half of the total costs of the project be raised from private sources.

\begin{figure}
\resizebox{\hsize}{!}{\includegraphics{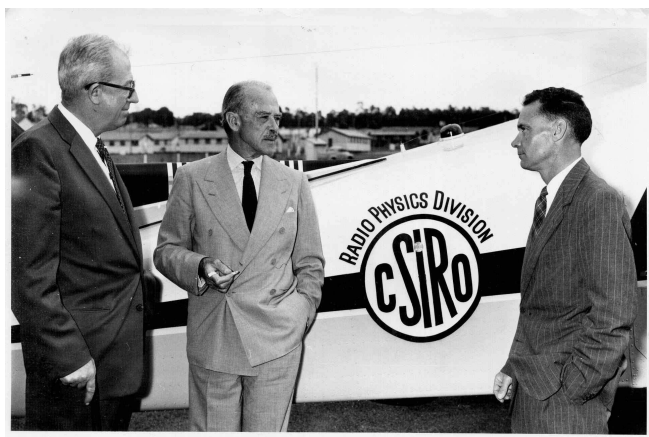}}
\caption{Finding the funds: CSIRO Chief Executive Fred White (left) and CSIRO Minister Richard Casey with Taffy Bowen in January 1958.  The light aircraft was used in other Radiophysics research programs such as air navigation and cloud physics.}
\end{figure}

Early in 1955 CSIRO established the Radio Astronomy Trust, with a board of trustees that consisted of a number of businessmen and pastoralists.  The Trust would have two roles, the first being to manage and invest the Carnegie funds until the construction of the GRT eventually got underway.  The second role was to seek out private sources of funding within Australia.  A range of wealthy individuals were approached but, apart from a few largely token donations, none were prepared to be the local version of an Andrew Carnegie.  Although some of the trustees themselves made generous donations, the fund raising attempts by the Radio Astronomy Trust were a disappointing failure.  The amount raised was less than 3\% of the final cost of the project.

By mid 1955 it was clear to Bowen that the funds promised would be nowhere near enough to build a GRT to rival the telescope already under construction at Jodrell Bank.  Once again he flew to the US to visit, cap in hand, a number of other philanthropic foundations.  At the Sloan and Ford Foundations he was informed that funding scientific research in another country halfway round the world was not part of the charter of either organisation.  His reception at the Rockefeller Foundation was however far more positive, helped no doubt by his wartime colleague Lee DuBridge who was an influential member of the Rockefeller Board of Trustees. In December 1955, the Rockefeller Foundation announced that, similar to Carnegie, it too would grant \$US 250,000 towards the cost of an Australian GRT.  With the Australian government's pledge to match the US funding, just over \pounds A 500,000 was now in hand, enough for the project to proceed.  

\begin{figure*}
\resizebox{\hsize}{!}{\includegraphics{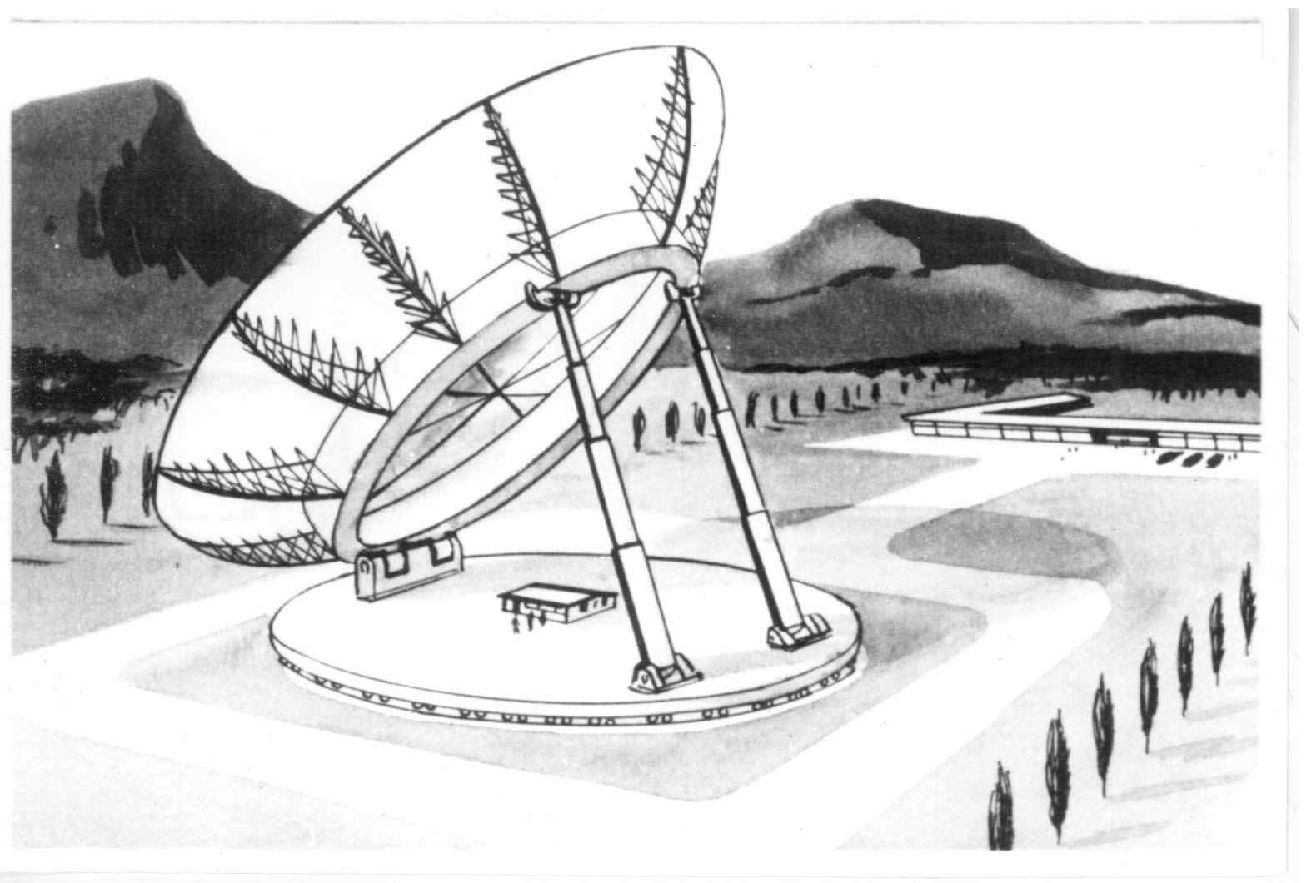}\includegraphics{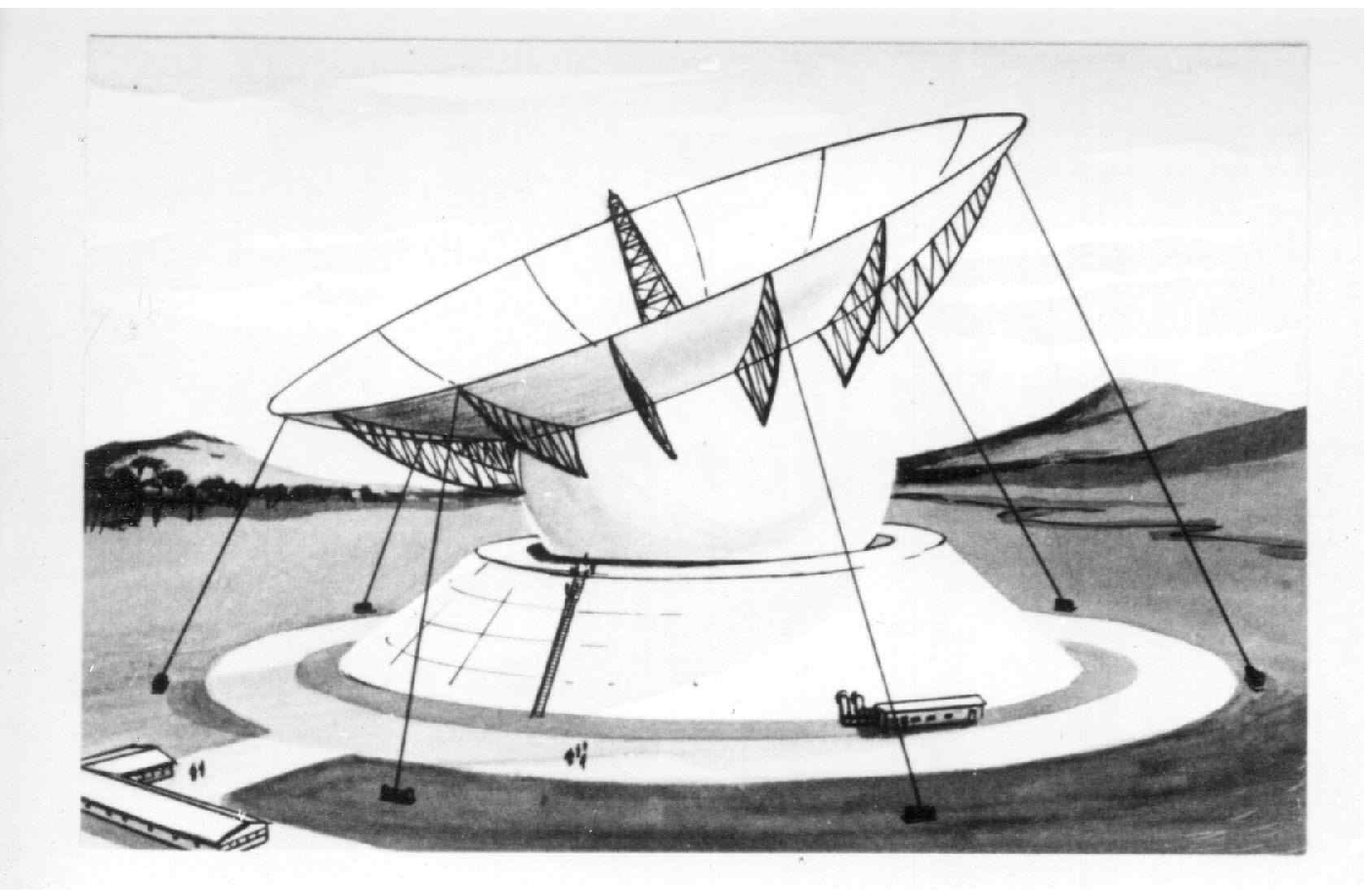}}
\caption{Two of the more unusual design concepts published in the 1955 publicity booklet:  (a) `jacks on turntable' where the dish is tilted by two jacks mounted on a rotating platform; (b) an `eyeball' design, submitted by a Chicago firm, where the dish is fixed to a section of a sphere which floats in a pool of water.}
\end{figure*}

\section{Designing a Giant Radio Telescope}

Immediately after the announcement of the Carnegie grant in May 1954, Bowen established a GRT planning committee consisting of senior Radiophysics staff and external engineering experts.  The committee's brief was to invite organisations to submit design concepts and then to assess their viability.  Various designs were received from a number of engineering firms in Australia and overseas, including one from the renowned American inventor Buckminster Fuller.  The progress of the GRT project was summarised in a publicity booklet (Radiophysics Lab 1955), which featured seven different design concepts (see Fig. 6).

For all the ingenuity of these designs, the real breakthrough in the design of the GRT came about by accident.  During a trip to London in 1955 Bowen was introduced to Barnes Wallis, the leading British inventor--engineer of his generation.  He was the designer of aircraft, such as the famous Wellington bomber which became the RAF's workhorse during WWII.  Wallis was best known for his invention of the bouncing `dambuster' bomb which destroyed a number of German dams:  the bomb would bounce across the surface of the water -- like a skipping stone -- hit the dam wall, sink to its base, and then explode.

Over lunch one day Bowen discussed with Wallis the GRT planned for Australia.  Wallis immediately came up with a few ideas and agreed to work on a design concept.  A couple of months later he proposed the design shown in Fig. 7.  Although it differs from the final Parkes design, it shows a number of important innovations.  One is the structure of the dish with its system of spiral and radial purlins. The design ensures that the dish surface maintains a rigid parabolic shape as it tilts from the zenith to the horizon and is subject to varying gravitational forces.  The second feature is known as the master equatorial, consisting of a small optical telescope, situated at the intersection of the two axes of rotation (altitude and azimuth), which can be pointed in a particular direction in the sky with great accuracy.  The dish is `slaved' to the master equatorial by a servo control mechanism and so achieves a high degree of pointing accuracy.

\begin{figure*}
\resizebox{\hsize}{!}{\includegraphics[width=8cm]{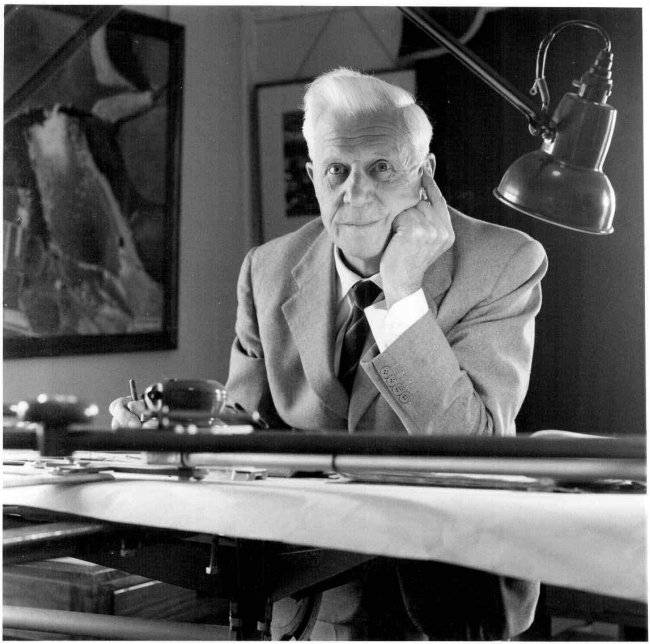}\includegraphics[width=8cm]{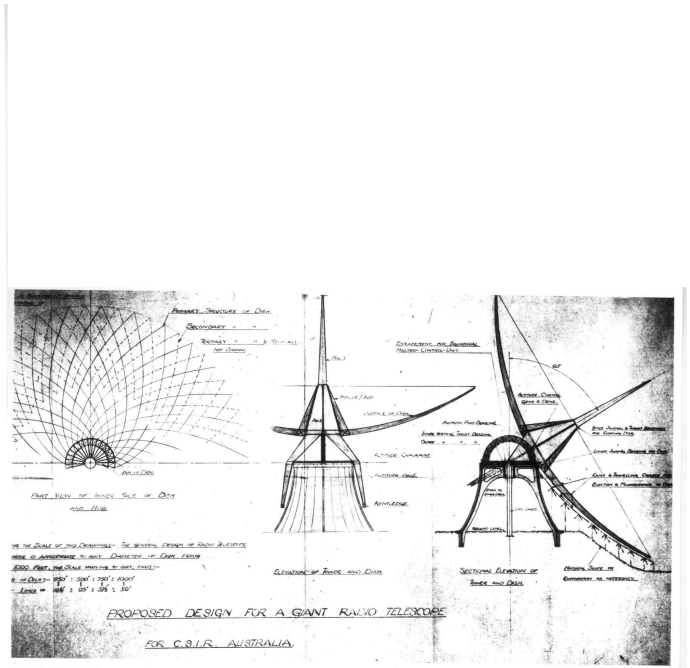}}
\caption{Barnes Wallis and his design concept for a GRT.  Wallis deliberately did not specify a diameter for the dish.  He believed that, depending on the funds available, his design could be scaled up to a diameter of 1000 ft (305 m). [courtesy: Vickers, London]}
\end{figure*}

After Wallis had produced his design concept, the plan was then to carry out the detailed design of the telescope in Australia, but it became clear that there were no local firms with the experience or expertise up to the challenge.  After several engineering firms in the UK were invited to bid, Freeman Fox \& Partners in London were chosen for the detailed design.  The firm specialised in the design of bridges, motorways and power stations.  In fact it was the firm's founder, Sir Ralph Freeman, who designed the Sydney Harbour Bridge, the most famous structure in Australia.  Senior partner Gilbert Roberts was appointed project leader and Harry Minnett from Radiophysics was seconded to the design team to work on the telescope's drive and control system and to provide the liaison between London and Sydney.

Throughout the planning of the GRT, Bowen and the Radiophysics group followed the progress of the telescope at Jodrell Bank with great interest.  Early in the project Lovell made a number of design changes in an attempt to improve its performance.  The major change, following the discovery of the hydrogen line in 1951, was the decision to operate down to a 21 cm wavelength, compared with the previous metre wavelength lower limit.  This meant finer, and heavier, mesh surface panels and the need to strengthen the support structure.  The various changes led to delays in the project, contractual problems with suppliers, allegations of a cover-up and, worst of all, a massive cost blow out.  Later, in his autobiography, Lovell (1990) recalled that he was probably fortunate he avoided going to prison over the whole affair!

The Australians learnt a lesson from Lovell's troubles.  Although Bowen frequently complained about slow progress at Freeman Fox, the design and construction of the Parkes Telescope went much more smoothly in comparison.  Jodrell Bank had been designed in relative haste and took over five years to build.  In contrast, Parkes took three years to design, but only two years to build.  A sketch of the final design is shown in Fig. 8.

\begin{figure}
\resizebox{\hsize}{!}{\includegraphics{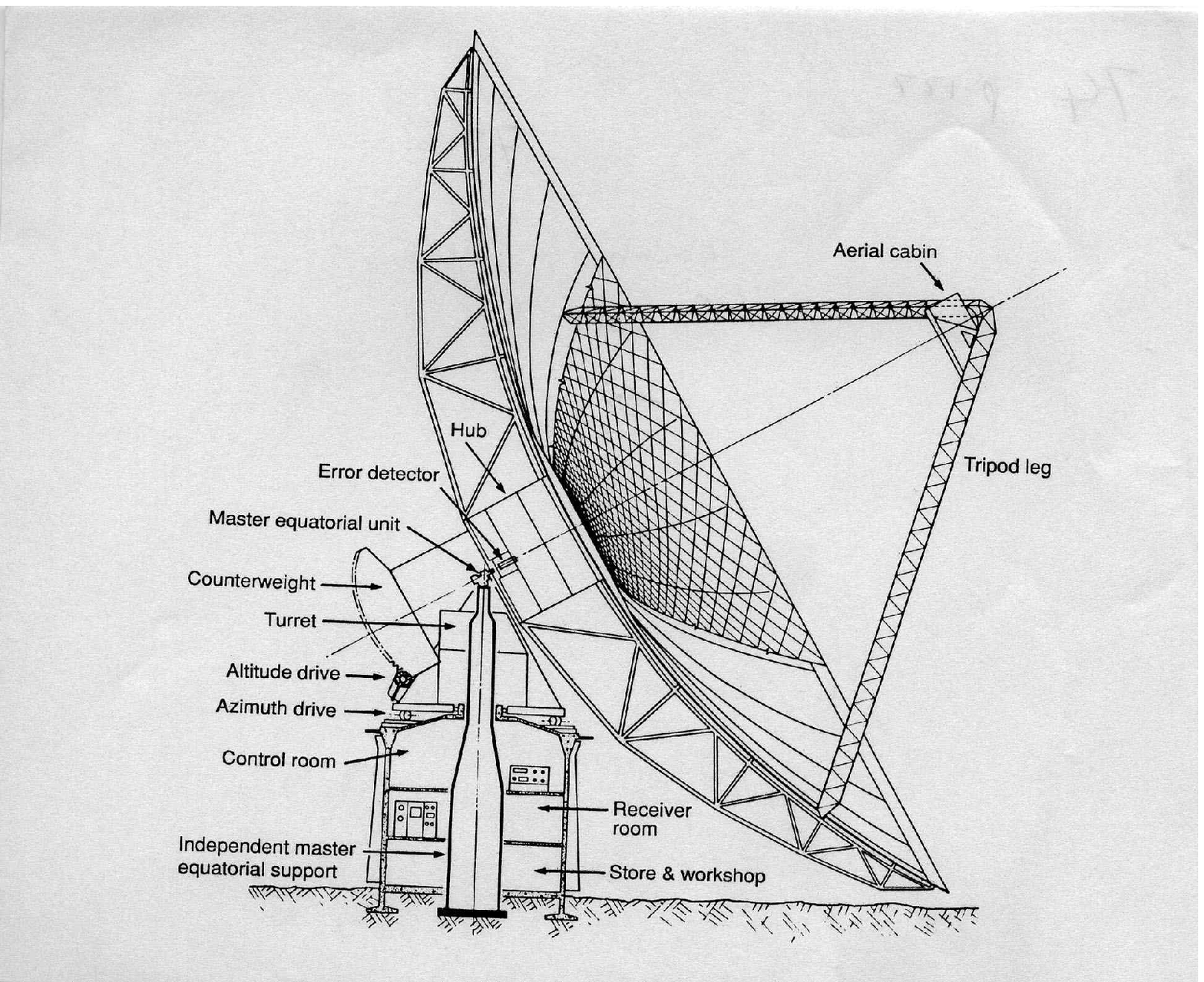}}
\caption{Major design features of the Australian GRT produced by Freeman Fox in April 1959.}
\end{figure}

\section{Search for a site}

The selection of the site near Parkes in central New South Wales for the new telescope took the best part of four years to settle.  Preliminary discussions on a suitable site began in mid-1954, prompted by press reports on the award of the Carnegie grant which saw a stream of letters arrive at Radiophysics from real estate agents, shire councils, chambers of commerce, and even a politician lobbying to have the telescope built in his electorate.  By early 1956 a list of over thirty possible locations had been compiled.

Several technical requirements were taken into consideration in shortlisting a site.  The ideal location would need to be geologically stable to provide a solid foundation capable of supporting a structure weighing close to 2000 tonnes.  The site would also need to have a mild climate free from ice and snow, with a low average windspeed all year round.  Above all, the site had to offer a very low level of radio interference.  To add to his troubles at Jodrell Bank, Lovell had been drawn into a long battle trying to persuade a local electricity authority to reroute a new high-voltage line away from the site. 

While Cheshire proved an uncomfortably noisy part of England, it was clear that a country as large and as sparsely populated as Australia would have no shortage of sites free of radio noise, and which would also meet the other geological and meteorological requirements.  Consequently, other logistical and political factors came into play.  Although for a time a site near Canberra was in the mix, the issue settled down into a two-way contest -- a site close to Sydney, or one `over the mountains' well to the west of Sydney.

On the grounds of convenience the ideal site would be not too distant from the centre of Sydney.  From the beginning the Radiophysics field stations had been no more than a couple of hours' drive away (see Fig. 1).  Staff had grown accustomed to a pattern of frequent, often daily visits to and from home to carry out their observing programs.  For these reasons most of the sites examined initially fell within a comfortable radius of Sydney, the best candidate to emerge being an area known as Cliffvale, at the foot of the Blue Mountains, 80 km south--west of Sydney.  Three sites in this area were selected and tests begun to gauge the levels of radio interference.

While the Cliffvale area became the favoured option, an extensive search was also made during 1957 for a location further west of Sydney, and west of the mountain range known as the Great Dividing Range.  Large areas in the sparsely populated region were inspected, with a site near Parkes chosen as the best of these `over the mountains' candidates.  Situated in the Goobang Valley, 20 km to the north of the Parkes township, this site was particularly well shielded from radio interference by a surrounding ring of hills.  Though not as accessible as Cliffvale the site, about 400 km west of Sydney, could be reached in five to six hours driving time or by a twice-daily air service to Parkes.

In March 1958 members of the radio astronomy group met to discuss the relative merits of the Cliffvale and Parkes sites.  Bowen did not attend the meeting as he wanted a democratic decision to be reached. Tests at both sites showed that Parkes had much lower noise levels than Cliffvale.  More importantly, with the projected industrial and residential expansion of Sydney to the west, the quality of the Cliffvale site would only deteriorate over the long term.  On these grounds Parkes had a clear advantage.  By the end of the staff meeting, the decision in favour of Parkes was unanimous.

\begin{figure}
\resizebox{\hsize}{!}{\includegraphics{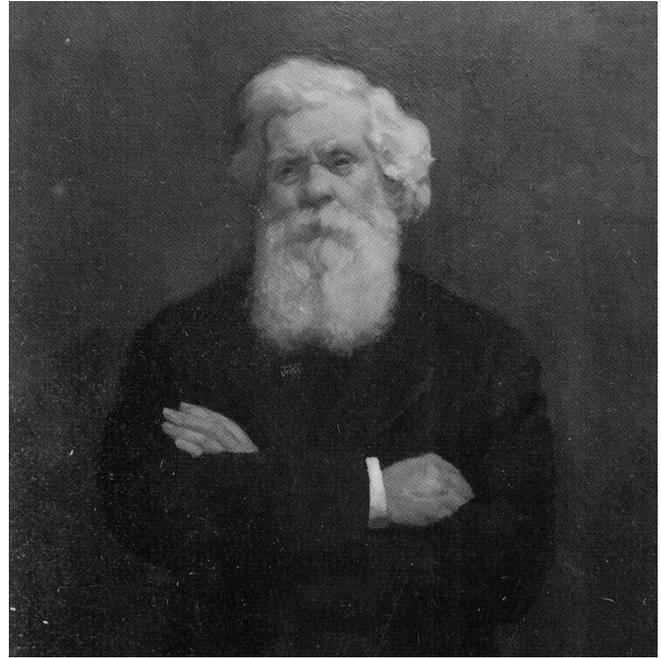}}
\caption{Sir Henry Parkes (1815–-96) is recognised as the founding father of Australian federation.  In 1873 he had a gold mining town named after him, and in 1961 a radio telescope. [portrait by Julian Ashton, Parliament House Collection, Canberra]}
\end{figure}

The Parkes district had first been explored by Europeans in the 1830s.  By the early 1860s settlers from the crowded coastal plains around Sydney had begun to move west and claim large tracts of land, dispossessing the indigenous people who had long occupied the open grasslands and rolling woodland.  In 1871 the population of the area accelerated sharply with the discovery of gold which saw thousands rush to pan the alluvial deposits in the creek beds.  In 1873 one of the colony's most prominent figures, Henry Parkes, toured the area ending the visit with an address to a large crowd gathered in the new goldfield town known as Bushmans.  To honour his visit, the people of Bushmans decided to change the name of their town to Parkes (see Fig. 9).

In March 1958 negotiations began with Austie Helm, the owner of a 360 hectare property known as Kildare in the Goobang Valley.  The area chosen initially consisted of a 50 hectare parcel but, to allow for the possible addition of other instruments in the future, two adjacent paddocks were included in the transaction to bring the total area to 170 hectares.  The negotiated price amounted to less than 2\% of the final cost of the project.

In the meantime various tests had been carried out at the site.  Test bores were sunk through the top soil which indicated a hard clay layer at a depth of about 12 metres.  A study of local meteorological records confirmed the area's favourable climate.  Although snowfalls, as well as wind gusts reaching 150 km per hour, had been recorded, the occurrence of such freakish weather was very uncommon.  Anemometers installed on the site showed that wind speeds were generally low, exceeding 30 km per hour only 1\% of the time.  Only on rare occasions would unfavourable weather shut down operations and force the telescope to be stowed in its safety position.

\section{Out to Tender}

After Freeman Fox had completed the design in April 1959, the next issue to be settled was who would build the telescope.  The choice of a suitable Australian company would have been ideal but, similar to the telescope design, there did not seem to any local firm with the necessary experience and expertise.  As a result, Freeman Fox decided to put the project out to international tender.  A compelling reason for this decision came with the launch of Sputnik in October 1957.  The birth of the space age created a sudden interest in large steerable dishes for tracking spacecraft.  Many companies particularly in the United States were eager to get in on the ground floor of this rapidly growing technology.  

After completion of the tender documents, Freeman Fox invited bids from a shortlist of companies that had already expressed an interest in the proposed Australian dish.  In view of the complex interplay between the structural, mechanical and control features of the telescope, bids were only accepted from companies prepared to take on the role of prime contractor.  Freeman Fox would retain responsibility for overall supervision of the construction, including inspection and testing of the multitude of components during manufacture.  By the close of the tender period in June 1959 bids had been received from seven companies, five from the US and one each from England and West Germany.  The lowest bid of \pounds A 611,000 came from the firm of Maschinenfabrik Augsburg Nurnberg (or MAN for short).  Most of the American firms priced themselves out of the project with bids exceeding \pounds 900,000.  Each of the seven companies quoted a completion time of under 27 months, with the MAN firm again leading the field with a promise of completion in the extraordinarily short period of 21 months.

With three major steelworks in West Germany, MAN had established itself as one of the world's leading producers of steel products.  About half the bridges constructed in Europe since the war had come from its workshops.  In July 1959 MAN was awarded the job as prime contractor, with three other firms as the principal subcontractors.  One was Associated Electrical Industries of Manchester, given responsibility for the servo-control system as well as all the cabling, lighting and communication systems for the telescope.  The other two were Askania-Werke of West Berlin, which would develop the master equatorial system, and Concrete Constructions of Sydney, which would erect the reinforced concrete tower.

Despite strong interest in the project by American firms, it is not surprising that MAN walked away with the contract.  With postwar recovery well advanced and West Germany's re-emergence as a major economic power, labour and production costs had remained low compared with other industrialised countries.  In contrast, the boom economy in the United States saw living standards jump well ahead of any other country.  Its high production costs made it difficult for American firms to compete openly on the international scene.  

Even though MAN's bid was 20\% lower than any other firm, it still created a financial problem for the project.  The \pounds 611,000 price tag applied only to the telescope itself, not to costs associated with on-site developments such as laboratory and residential buildings, roads, landscaping and the provision of essential services.  Once these additional outlays were included the total cost of the project climbed to almost \pounds 800,000.  In contrast, the actual funds in hand by mid 1959 totalled only \pounds 560,000, consisting of the Carnegie and Rockefeller grants made in 1954 and 1955, the modest sum collected by the Radio Astronomy Trust, plus the government pledge to match these funds pound-for-pound.

With the project facing a substantial shortfall, Bowen began another round of visits to the American foundations.  The Carnegie Corporation refused to supplement its original grant, but the Rockefeller Foundation provided a sympathetic ear and promised a further \pounds 50,000.  Following lengthy negotiations in Canberra, the government agreed in January 1960 to cover the remaining shortfall so that, at last, the final piece of the financial jig-saw had fallen into place.  Although the cost overrun was not insignificant, it was relatively modest for a project developing a brand-new technology, a credit to how well Freeman Fox had carried out the design.

\begin{figure*}
\centering{
\resizebox{\hsize}{!}{\includegraphics[width=6cm]{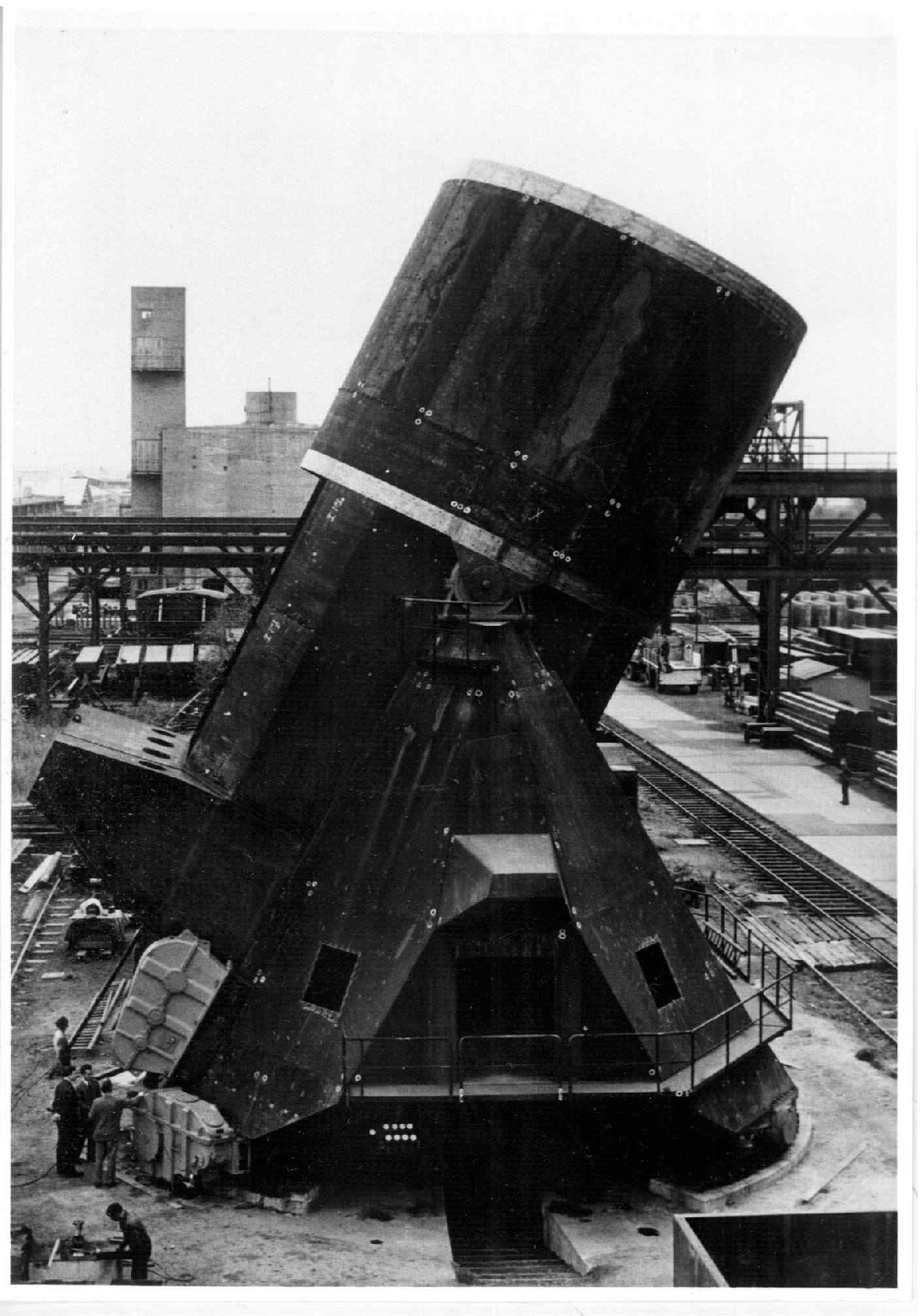}\includegraphics[width=8cm]{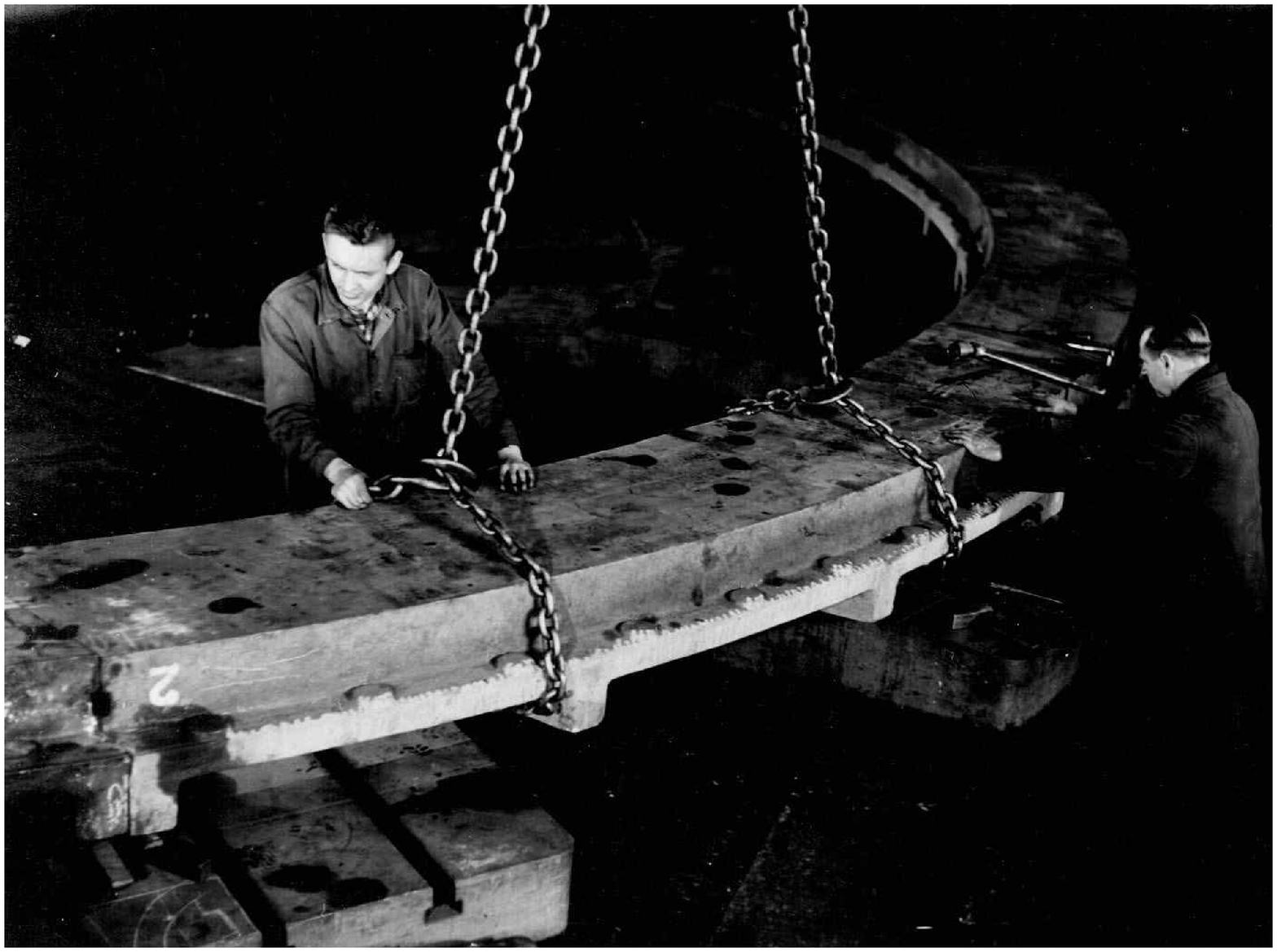}}}
\caption{(Left) MAN carried out a trial assembly of the telescope’s turret and hub at its Gustavsburg plant near Frankfurt.  (Right) Casting the steel roller track which sits on top of the telescope tower and supports the massive superstructure of the dish.}
\end{figure*}

\begin{figure}
\resizebox{\hsize}{!}{\includegraphics{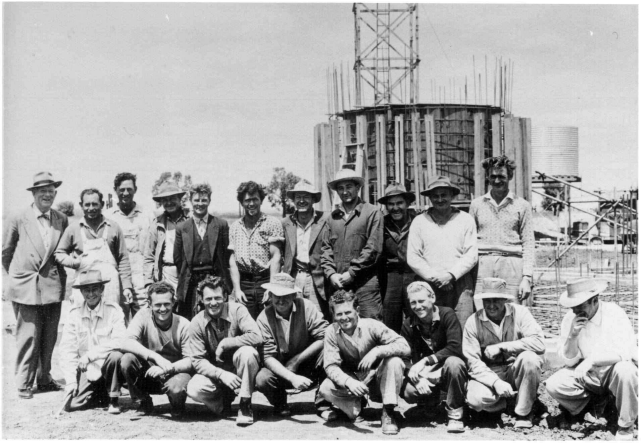}}
\caption{The tower was constructed by the Sydney firm Concrete Constructions which recruited most of the tradesmen from the Parkes district.}
\end{figure}

\section{Construction of the Dish}

By September 1959 the construction of the telescope had got off to a flying start -- not only at Parkes but also at three different centres halfway round the globe.  Associated Electrical Industries began on the servo-control system in Manchester, Askania-Werke started manufacture of the master equatorial unit in West Berlin, and MAN began casting some of the massive steel components at its Gustavsburg plant.  By May 1960 the fabrication of the turret, hub and counterweight structures had been completed.  A trial assembly of these components was then made on a section of the steel roller track so that various tests and measurements could be made (see Fig. 10).  After these trials the assembly was dismantled and packed into crates ready for shipment to Australia.

In the meantime the Sydney firm, Concrete Constructions, had completed the excavations for the telescope's foundations and in November 1959 began erection of the reinforced concrete tower (Fig. 11).  By March 1960 the last concrete had been poured.  The tower was then allowed to stand for several months so that any settlement of the foundations would occur well before cranes began lifting the telescope's 1000 tonnes of steel superstructure into place.

The 170 hectare site also underwent a transformation, beginning with the construction of an access road from the main Newell highway and an internal network of roads.  Plans were also made for the construction of a cluster of buildings situated a kilometre north of the telescope.  This `village' area would consist of two residential houses for engineering staff, as well as a larger dormitory building for visiting research and technical staff.  A combined garage and fuel store was also erected for housing site vehicles, a tractor and other essential items such as firefighting equipment.

In September 1960 the MAN construction crew arrived on site, together with some of the steel components cast in West Germany and brought from Sydney by rail and road.  The first task involved fitting the circular steel roller track to the top of the tower.  Once the track had been bolted into place, precision measurements showed that over the complete 35 metre circumference of track, no part deviated from a perfectly flat surface by more than 0.25 mm.  Then the MAN crew retraced the steps followed at the Gustavsburg workshop and began to assemble the steelwork to support and steer the dish structure.

First came the turret, a compact and very rigid structure that rotates around the azimuth roller track on four sets of bogies.  Next the cylindrical hub was mounted on two strong A-shaped steel frames of the turret.  The hub supports the dish structure and movement in altitude is carried out by a rack and pinion drive on each side of the hub.  These racks are in turn attached to heavy counterweight boxes filled with 400 tonnes of concrete ballast mixed with steel punchings, counterbalancing the weight of the dish itself.  Slightly more counterweight was added than actually needed to balance the dish, so that the dish would always have the tendency to return to its upright or zenith direction.

\begin{figure*}
\centering{
\resizebox{\hsize}{!}{\includegraphics[width=8cm]{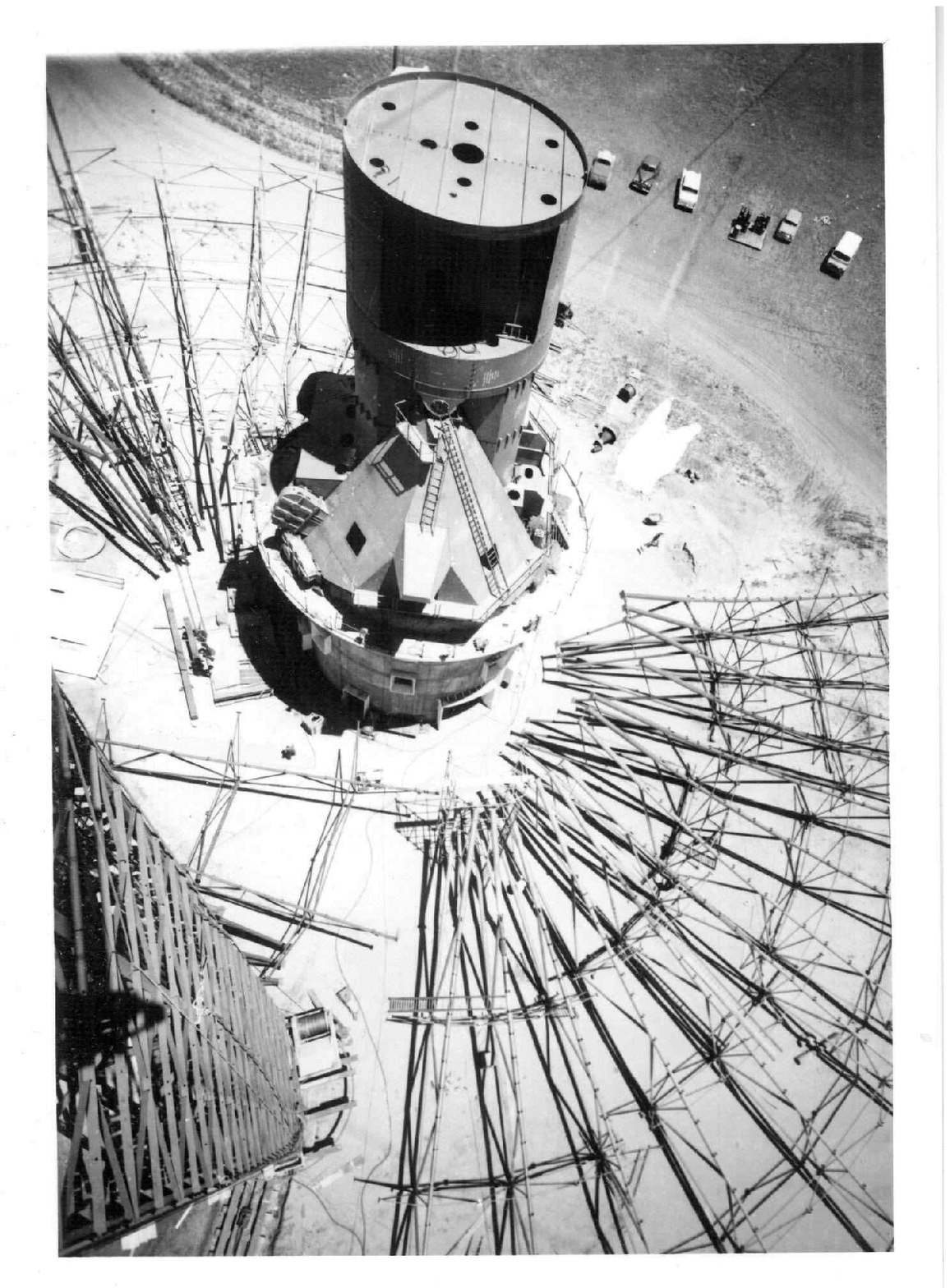}\includegraphics[width=8cm]{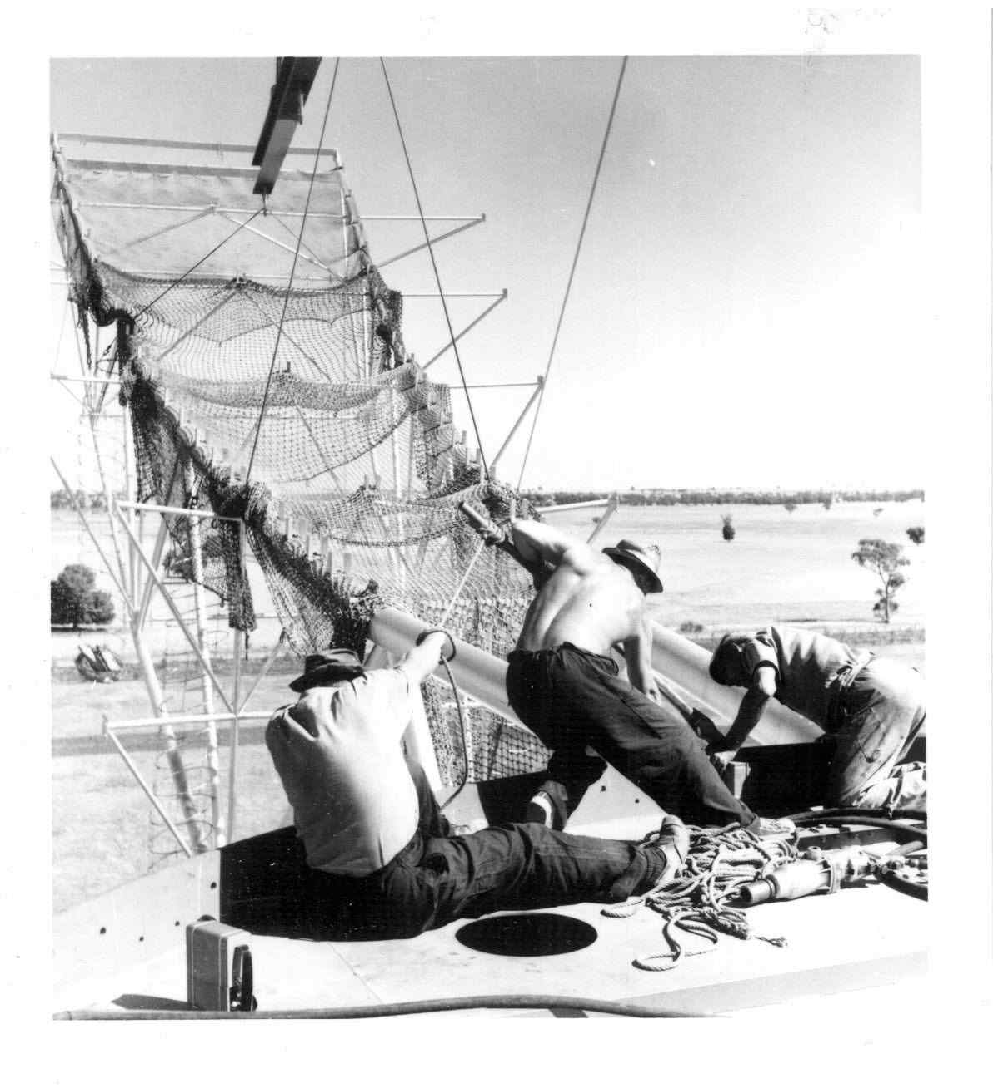}}}
\caption{(Left) The rib sections of the dish were fabricated on site and welded at night to minimise thermal stresses.  A thick layer of sand litters the site, used to sandblast the ribs prior to painting.  (Right) Each rib section was lifted into place and bolted to the hub. }
\end{figure*}

With the last of the counterweight concrete poured by March 1961, work began on the erection of the dish structure.  Fabrication of the thirty radial ribs to form the dish had already been underway at ground level (Fig. 12).  By day the tubular steel was cut to length and placed in jigs set in massive blocks of concrete.  Then, at night, welding of the joints took place when all the steel components were at a uniform temperature, minimising the possibility of locking thermal stresses into the structure.

After sandblasting and painting, these rib structures were then connected in twos or threes, and each of these `slices of pie' was lifted into place, bolted to the hub and then fixed to its neighbouring segments by a series of ring girders.  Next came a period of comparative lull as each rib and ring girder was carefully surveyed and adjusted.  Work then began on attaching the two sets of steel cross members or purlins, each shaped in the form of an extended spiral, but opening out in opposite directions. 

In the meantime the aerial cabin had been lifted into position.  Supported by three legs and set slightly above the focal point of the parabolic dish, the cabin provided space for receiver equipment and enough room for two people to work in relative comfort.  Access to the cabin is provided by a small elevator in one of the tripod legs or by ladders in the other two.

With the framework of the dish largely complete, work began laying the reflecting surface.  First, segments of thick steel plate were used to form the central section of the surface (17 m in diameter), the plates welded together to provide the dish with extra stiffness.  The remaining part of the surface was then covered with over one thousand mesh panels, each woven from high-tensile steel wire.  The panels were laid out ring by ring, working outwards, with each panel fastened to its purlin support (Fig. 13).

\begin{figure}
\resizebox{\hsize}{!}{\includegraphics{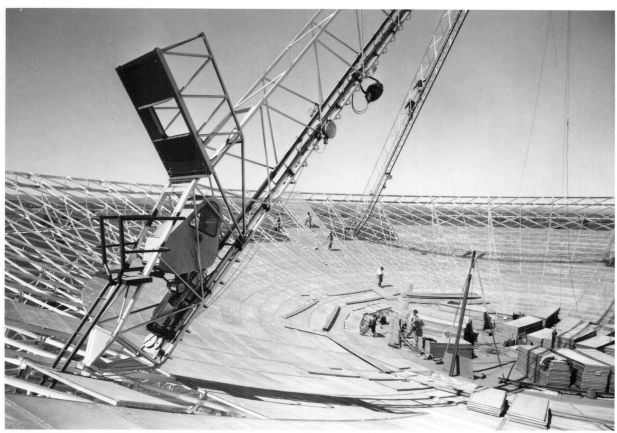}}
\caption{Over 1000 steel mesh panels were fitted to form the reflecting surface of the dish.  The aerial cabin above can be reached via the small lift within the tripod leg.}
\end{figure}

In parallel, the construction crew carried out a number of finishing touches on the rest of the telescope.  By the end of August 1961 the structural work was virtually complete, just on two years after construction began and only slightly over the ambitious 21 month schedule MAN had promised at the outset.  Askania was the only contractor unable to keep up the pace, having fallen three months behind early in the manufacture of the master equatorial unit and, despite constant pressure applied by MAN, unable to make up the time.  Yet, by any standards, the construction was proceeding remarkably smoothly with few problems or delays.

The weeks leading up to the inauguration in late October became a period of feverish activity for the teams of technicians, electricians and painters.  To avoid congestion, the day was divided into a day shift, followed by a night shift lasting until midnight.  There was an abundance of wildlife at the site, some of it unwelcome.  Spiders and snakes were a constant danger, while at night the floodlights attracted vast swarms of insects.

By early October the master equatorial had been installed and wired to the control desk in the telescope's tower.  Meanwhile, despite problems with the control desk, test drives of the telescope in azimuth had gone ahead.  With only a subdued hum from the motors, the motion of the telescope's superstructure proved impressively smooth.  By mid October, the Parkes telescope was finally ready to tip in elevation -- an occasion that marked the culmination of six years of planning and construction.

\section{Concluding remarks}

The official opening of the Parkes telescope by the Australian Governor-General, Lord De L'Isle, on Tuesday, 31 October 1961 marked a special day for science in Australia.  An idea first conceived ten years earlier, the telescope had taken just over three years to design and a further two years to construct.  From the outset the project had been coordinated on an international level.  A substantial part of the finance had been provided by American philanthropic foundations, the design had been carried out by British engineers, the construction by German steelmakers and the overall coordination of the project directed by CSIRO's Radiophysics Laboratory.

Almost 500 official guests gathered at the site for the inauguration.  Some made the long drive from Sydney, Canberra or Melbourne, but most arrived on chartered flights.  For several hours the small Parkes airstrip resembled a major terminal with aircraft arriving every few minutes.  The guest list read like a `who's who' of Australian science, and also included a large number of politicians, senior public servants and representatives of the companies that had designed and built the telescope. Before the ceremony commenced the visitors inspected the telescope, clambering along gangways and out on the catwalks directly beneath the surface of the dish.  By the time the vice-regal party arrived, the guests who had already lunched in marquees pitched near the telescope sat waiting in front of the official dais.

\begin{figure}
\resizebox{\hsize}{!}{\includegraphics{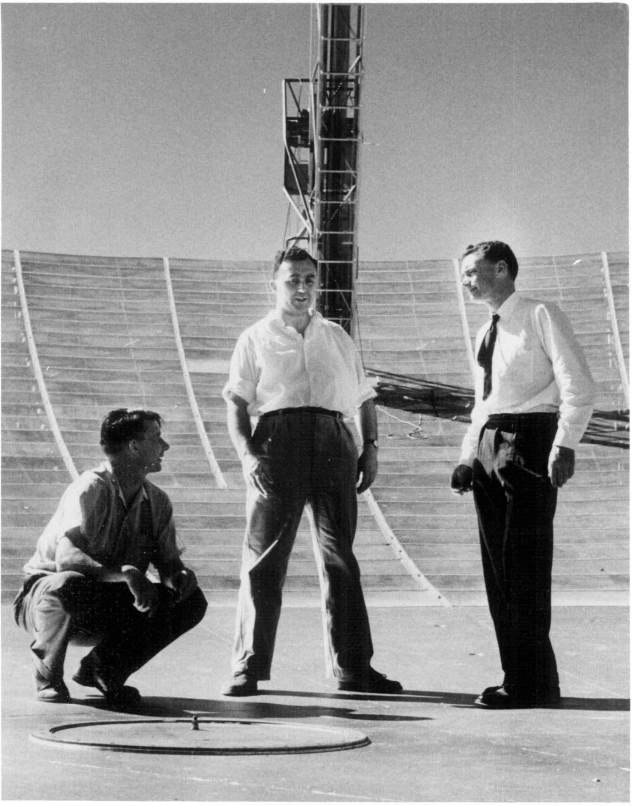}}
\caption{Three key players:  John Bolton, the inaugural director (left), Mike Jeffery, the Freeman Fox project manager, and Taffy Bowen.}
\end{figure}

Unfortunately, the day did not quite go to plan.  The weather conditions could hardly have been worse.  The temperature climbed into the mid-thirties and gale-force winds gusting up to 80 km per hour lashed the site, whipping up thick clouds of red dust and drowning out the sound from the public address system.  The irony was, of course, that one reason for selecting the Parkes site was its excellent record of low wind speeds!

After the speeches, the crowning moment planned was for the Governor-General to press a button which would slowly tilt the telescope's 500 tonne dish from the upright position down to its lowest point facing the assembled guests.  However, the strong gusty winds far exceeded the design safety limit and driving the telescope would have involved a high risk of serious damage.  Instead, in a last minute arrangement, two of the Parkes staff bravely made their way up to the aerial cabin high above the centre of the dish and unfurled a large Australian flag.  The wind tore at the flag, holding it in an almost perfectly extended position.  The Parkes dish was now officially opened.

\begin{figure*}
\resizebox{\hsize}{!}{\includegraphics{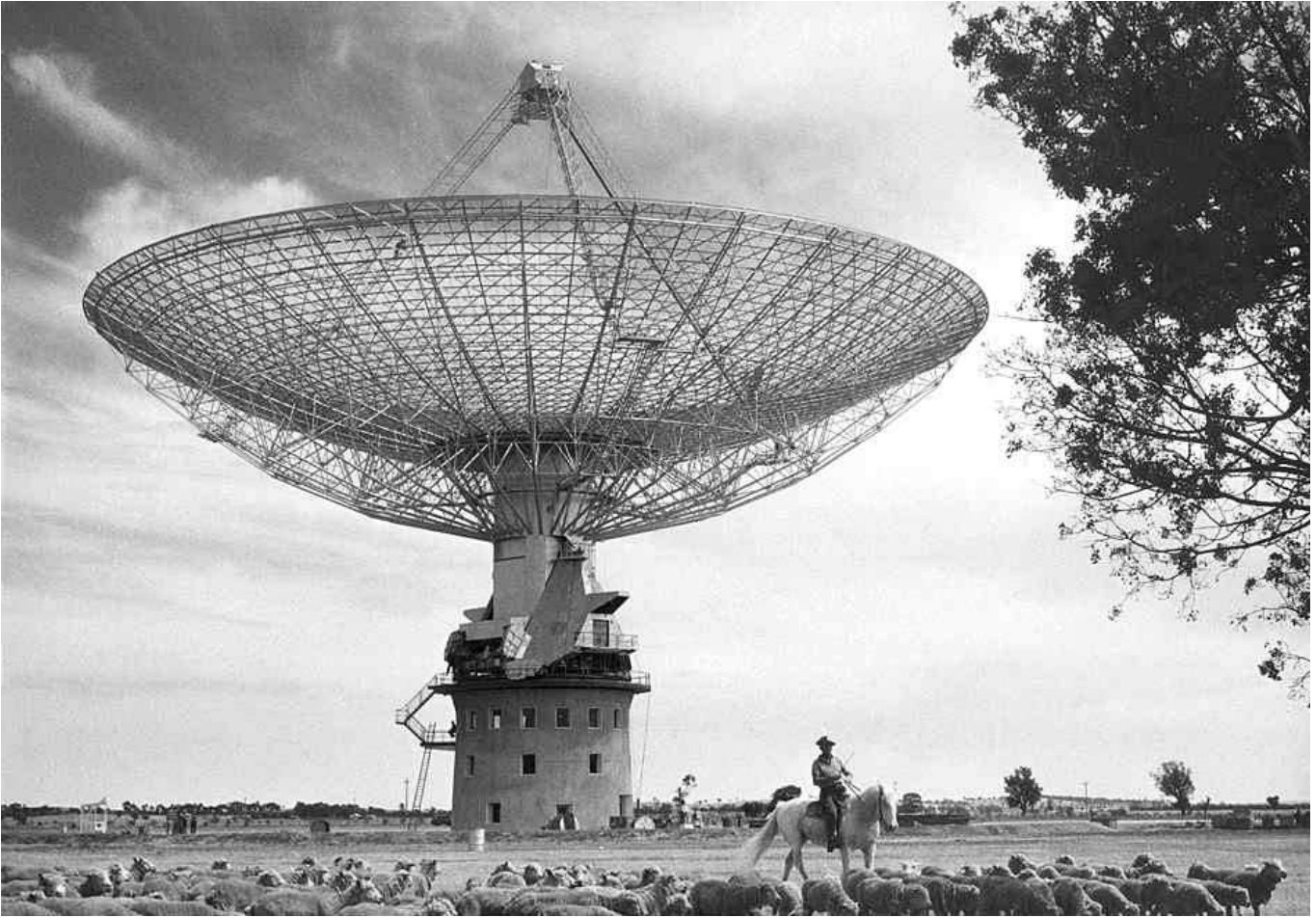}}
\caption{The Parkes dish shortly before its inauguration on 31 October 1961.  The horseman is `Austie' Helm who sold part of his farm to CSIRO to provide a site for the telescope.  Later, the site was named the Australian National Radio Astronomy Observatory (ANRAO). [courtesy: National Library of Australia] }
\end{figure*}

The inauguration was an important event in the development of Australian science.  On a purely scientific level, the telescope provided Australian astronomers with the most powerful and versatile instrument of its type in the world, one which immediately produced a stream of significant and, at times, fundamental discoveries.  On a different level, the Parkes telescope also had a major impact in shaping the way astronomy developed in Australia.  In contrast to the 1950s, when small teams at Radiophysics built and had exclusive use of their own instruments, Parkes would operate as an observatory and have more in common with the great optical telescopes of the world.  A fulltime specialist group of technicians would look after the maintenance and routine operation of the telescope and radio astronomers would now have to compete with their peers for observing time.  A new breed of radio astronomer emerged and also a new way of doing radio astronomy.  In this respect the completion of the Parkes telescope marked the arrival of radio astronomy in Australia as a mature scientific discipline.  In some respects too, it marked the end of the innovative, colourful and jack-of-all-trades period in radio astronomy.

The Parkes telescope also had a major impact on the broader development of astronomy in Australia.  With the massive investment of staff and funds in Parkes, only limited resources remained to support the other research programs which had flourished at Radiophysics in the 1950s.  This heavy commitment to Parkes sparked a major conflict for these remaining resources. Although we cannot explore this further here, this conflict eventually led to the departure of several leading members of the Radiophysics group to continue their careers in astronomy elsewhere in Australia and overseas, principally the United States.

Taffy Bowen and the engineers at Freeman Fox probably expected a productive lifetime for the Parkes dish of about twenty years and, with wishful thinking, they may have even hoped for thirty years.  Given that the lifetime of the various Radiophysics instruments built in the 1940s and 1950s was a few years at best, twenty years or more would have seemed a long time, an excellent return on investment.  However, they could not have foreseen the extraordinary developments that were to occur in areas such as receiver technology and computing.  Similarly, they could not have foreseen that the superb design would allow ongoing improvements to the structure of the telescope and to the accuracy of its surface.  Today, as it enters its sixth decade, the Parkes telescope is estimated to be ten thousand times more sensitive than it was in 1961 -- a wonderful testimony to Taffy Bowen and his vision.

\vskip0.2cm

This article is based on a talk presented at the Parkes Open Days on 8--9 October and at the Parkes@50 Symposium on 31 October 2011.  I am grateful to CSIRO Astronomy and Space Sciences (CASS) for providing travel funding to attend both events. I am also grateful to Professor Ron Ekers for his continuing support of this historical research.  Unless indicated otherwise, all images in this article are reproduced with the permission of the CASS Photo Archives.

\end{document}